\documentclass[a4paper,11pt]{article}

\usepackage[numbers,sort&compress]{natbib}
\usepackage{amssymb}
\usepackage{xcolor}
\usepackage{fancyhdr}
\usepackage{sectsty}
\usepackage{graphicx}
\usepackage{url}
\usepackage{hyperref}

\fancypagestyle{plain}{%
  \fancyhf{}\fancyfoot[R]{Page \thepage}%
}

\newcommand{\apj}{ApJ}
\newcommand{\apjl}{ApJL}

\newcommand{\nat}{Nature}

\newcommand{\aap}{A\&A}

\begin{document}

\title{
\vspace{-2.5cm} {\em 
{\bf \Large Traversing the Galactic Centre in space and time}}}


  \author{{\bf Michal Zaja\v{c}ek}$^{1*}$, {\bf Bo\.{z}ena Czerny}$^2$,  {\bf Martin Mondek}$^1$, {\bf Samik Mitra}$^3$, {\bf Matúš Labaj}$^1$,\\{\bf Tomáš Ondro}$^4$, {\bf Jan Jan\'{i}k}$^1$, {\bf Ji\v{r}\'{i} Du\v{s}ek}$^5$
 }

\date{}
\maketitle

$^1$ Department of Theoretical Physics and Astrophysics, Masaryk University, Kotl\'{a}\v{r}sk\'{a} 267/2, 611 37 Brno, Czech Republic; email: \href{zajacek@mail.muni.cz}{zajacek@mail.muni.cz}

$^2$ Center for Theoretical Physics, Polish Academy of Sciences, Al. Lotnik\'{o}w 32/46, 02-668 Warsaw, Poland

$^3$ International Center for Theoretical Sciences, Tata Institute of Fundamental Research, Bangalore, India

$^4$ Department of Technology and Automobile Transport, Faculty of AgriSciences, Mendel University in Brno, Zem\v{e}d\v{e}lsk\'{a} 1, 61300 Brno, Czech Republic 

$^5$ Brno Observatory and Planetarium, Krav\'{i} hora 2,
616 00 Brno, Czech Republic

\begin{quote}
\noindent
At IAU Symposium 405 in Brno, the Galactic Center emerged as a nearby laboratory for galactic nuclei, linking the central supermassive black hole, dense stellar populations and multiphase gas into a connected ecosystem across space and time.
\end{quote}

The Galactic Center is often identified with its central supermassive black hole, Sgr~A*. Yet the black hole governs gravitationally only the innermost few parsecs of the Milky Way, while the surrounding Nuclear Star Cluster, Nuclear Stellar Disc and Central Molecular Zone (CMZ) shape the dynamics of stars and gas on progressively larger scales. Understanding how these components interact is essential not only for reconstructing the history of our own Galaxy, but also for interpreting galactic nuclei more generally.

\begin{figure}
    \centering
    \includegraphics[width=0.9\textwidth]{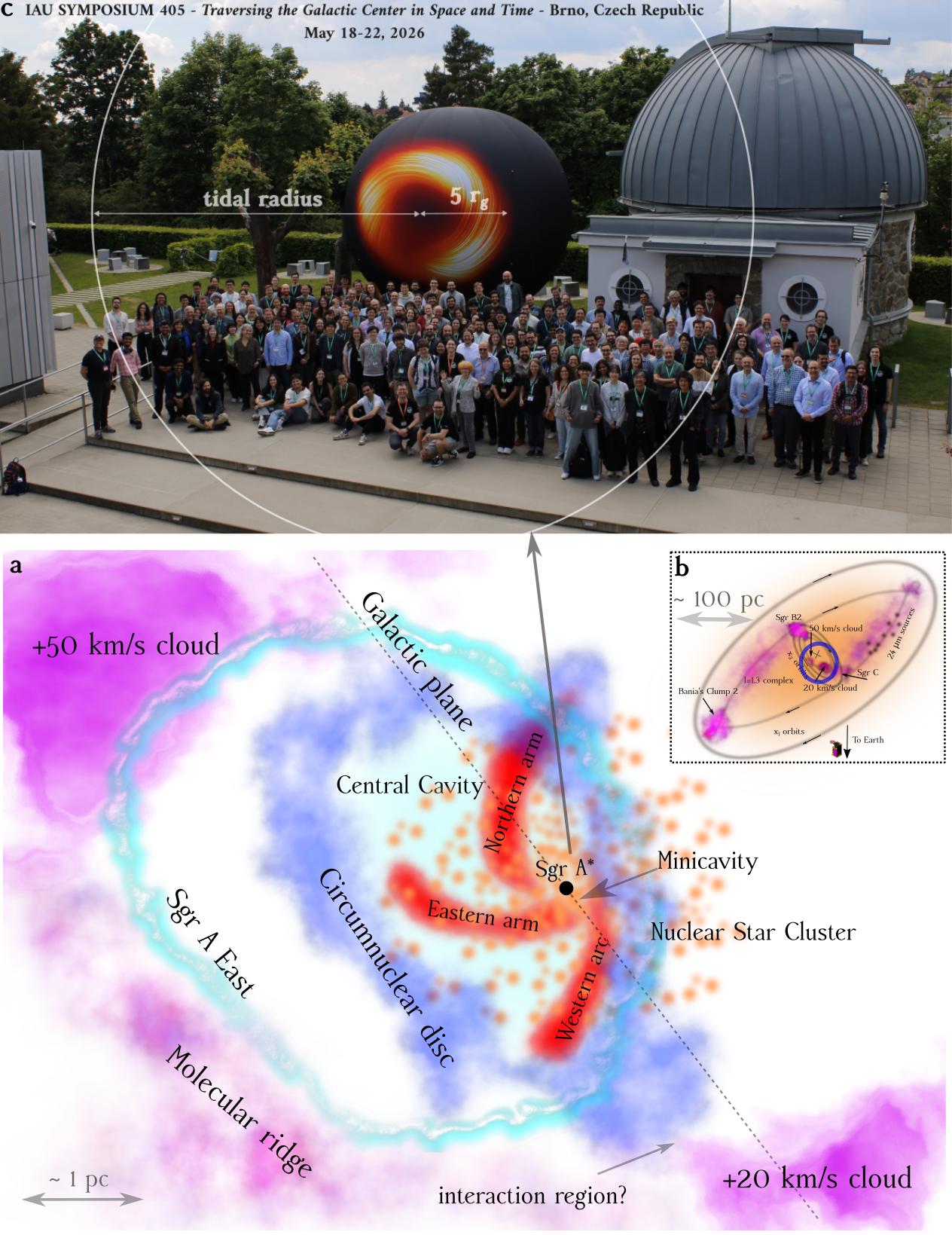}
    \caption{{\bf Complex dynamics in the Galactic Center.} Panel a) shows the innermost region of the CMZ, including the non-thermal supernova remnant Sgr~A East, the thermal Sgr~A West with the three Minispiral streamers, the Nuclear Star Cluster, the Central Cavity, and the Circumnuclear disc at $\sim2$--$5\,{\rm pc}$ from Sgr~A*. Panel b) (top right inset) illustrates the innermost Galactic bar, where gas is channelled inward along the dust lanes ($x_1$ orbits) into the $\sim100$ pc CMZ on $x_2$ orbits, highlighting the main molecular complexes (Sgr~B1, Sgr~B2, 20 and 50 ${\rm km\,s^{-1}}$ clouds, and Sgr~C). The blue circle marks the $\sim10$ pc region shown in Panel a. Panel c) shows the participants of IAU Symposium 405 with the 6-meter black-hole model displaying the polarized \textit{Event Horizon Telescope} image of Sgr~A* (photo credit: Lea Szakszonová).}        \label{fig_CMZ_SgrA}
\end{figure}

These themes were at the heart of the International Astronomical Union Symposium 405, \href{https://gc2026.muni.cz/}{``Traversing the Galactic Center in Space and Time''}, the 15th edition of the Galactic Center workshop series \cite{2026GCNew..29a...2Z}. More than 200 researchers gathered in Brno, Czech Republic, to discuss observational and theoretical advances spanning the event horizon of Sgr~A* to large molecular complexes in the CMZ; see Fig.~\ref{fig_CMZ_SgrA} for the multiscale Galactic Center region illustration including the participant group photo. A recurring message throughout the meeting was that the Galactic Center should no longer be viewed as a collection of individual phenomena, but as an interconnected ecosystem in which black-hole accretion, star formation, stellar feedback, gas dynamics and magnetic fields continuously influence one another.

\section*{From the Central Molecular Zone to the event horizon}

One of the central questions discussed at the symposium concerned the striking contrast between the vast molecular gas reservoir of the CMZ and the extremely low luminosity of Sgr~A* (talks by Steven Longmore and Mark Morris). Although the Galactic bar efficiently channels gas towards the inner Galaxy, only a tiny fraction ultimately reaches the black hole. Instead, much of the inflowing material is converted into stars, dispersed by stellar winds and supernova explosions, or expelled through outflows from the hot accretion flow surrounding Sgr~A* \cite{2013Sci...341..981W}. The Galactic Center therefore appears to regulate itself through feedback operating across many spatial scales (Zixuan Feng).

This perspective also reshapes our understanding of star formation. The CMZ provides a nearby laboratory for studying star formation under conditions resembling those in high-redshift starburst galaxies, with unusually high gas densities, turbulence and magnetic fields. Yet fundamental questions remain unresolved. Young massive stars are observed within the central parsec (Zoë Haggard) despite the intense tidal field of Sgr~A* and the extreme physical conditions of the surrounding circumnuclear medium \cite[the ``paradox of youth'';][]{2003ApJ...586L.127G}. Furthermore, the evolutionary connection between the Nuclear Star Cluster and the Nuclear Stellar Disc, which are cospatial with the gaseous circumnuclear disc and the CMZ, respectively, remains uncertain, though there are observational and theoretical indications for the ``inside-out'' growth (Mattia Carlo Sormani, Nils Ryde)
. However, some of the long-term mysteries are being solved, for instance new Atacama Millimeter/submillimeter Array observations (Steven Longmore, Laura Colzi) can trace the transition from supersonic to subsonic gas motions \cite{2026arXiv260220340L} and thus reveal where star formation becomes possible in this extreme environment.

\section*{From quiescence to an active nucleus and back again}

Although Sgr~A* is presently one of the faintest known supermassive black holes, it remains highly variable (talks by Sebastiano von Fellenberg and Farhad Yusef-Zadeh). Near-infrared and X-ray flares, now associated with orbiting hot spots \cite{2018A&A...618L..10G}, probe plasma properties within only a few gravitational radii of the event horizon, although their physical origin---magnetic reconnection or outflowing plasma---is still debated \cite{2018A&A...618L..10G,2025A&A...696A..10A} (Braden Seefeldt-Gail). These observations provide an exceptional opportunity to test models of black-hole accretion in the low-luminosity regime.

At the same time, multiple lines of evidence demonstrate that the current quiescent state is only temporary. X-ray echoes reflected by molecular clouds reveal that Sgr~A* experienced powerful outbursts only a few centuries ago (Maïca Clavel, Frederic Marin, \cite{2023Natur.619...41M}), while the Fermi and eROSITA bubbles indicate much stronger activity several million years ago (Fulai Guo, \cite{2019Natur.567..347P}). Whether these episodes were driven by enhanced accretion, nuclear starbursts or both \cite{2012ApJ...756..181G}, they reinforce a central conclusion of the symposium that emerged from the talks in the sessions focusing on both the Sgr~A* immediate environment and the larger-scale CMZ: the Galactic Center is a dynamic system that repeatedly cycles between quiescent and active phases as gas, stars and the central black hole evolve together. 

\section*{The Enduring Legacy of Brno Scholars in modern Galactic Center Research}

The symposium also highlighted Brno's scientific heritage and its continuing relevance to Galactic Center research. Alongside Gregor Johann Mendel, founder of genetics, participants reflected on two Brno-born scholars whose ideas still resonate with modern astrophysics.

Ernst Mach (1838–1916) influenced Einstein's development of general relativity through his ideas on inertia, while his studies of supersonic motion remain directly relevant to Galactic Center physics. Supersonic stellar winds generate bow shocks around stars  \cite{2014A&A...567A..21S}, and the transition from supersonic to subsonic turbulence in the Central Molecular Zone appears closely linked to the onset of gravitational collapse and star formation.

The meeting also marked the 120th anniversary of Kurt Gödel (1906–1978). Inspired by his incompleteness theorems \cite{2026GCNew..29e...1P}, Steven Longmore jokingly introduced a "Gödel–Schödel scale" to describe how (in)completely individual Galactic Center problems are understood. The humour reflected a broader message: despite remarkable observational progress, many fundamental questions remain open.

\section*{What is next for the Galactic Center}

The meeting helped identify several major challenges for the coming decade (Mark Morris). These include the origin of the young stars in the central parsec, possible ongoing star formation near Sgr A*, the manifestation of nuclear outflows, the demographics of stellar remnants and intermediate-mass black holes, the apparent scarcity of pulsars, and the role of magnetic fields in linking horizon-scale accretion to Galactic-scale structures.

Addressing these questions will rely on increasingly multi-wavelength and multi-messenger observations, together with next-generation facilities such as the Extremely Large Telescope, the Square Kilometre Array Observatory \cite{2026arXiv260625051S}, and continued observations with the James Webb Space Telescope \cite{2023arXiv231011912S}.

More broadly, the symposium underscored a conceptual shift in the field. Rather than treating Sgr A*, the stellar populations and the CMZ as separate systems, researchers increasingly view the Galactic Center as an interconnected ecosystem in which gas inflow, star formation, stellar feedback, magnetic fields and black-hole accretion regulate one another across many orders of magnitude in scale and time. As new observational facilities extend our view of the Galactic nucleus, they are expected to resolve long-standing questions while uncovering entirely new ones.

\section*{Acknowledgements}
The organizers of the Galactic Center workshop 2026 (IAU Symposium 405) thank the Grant Agency of the Czech Republic for the financial support via the Junior Star grant no. GM24-10599M  ``Stars in galactic nuclei: interrelation with massive black holes''. We also thank the Brno Observatory and Planetarium and its employees for providing the venue and the technical equipment. We are grateful to the students and volunteers from the Faculty of Science, Masaryk University, for much appreciated help during the symposium.

\section*{Competing interests}
The authors declare no competing interests.

\section*{Corresponding author}

Correspondence to \href{zajacek@mail.muni.cz}{Michal Zaja\v{c}ek}. 



\vspace{-0.4cm}


\end{document}